\documentstyle[11pt,moriond,epsfig]{article}
\def\jour#1#2#3#4{{#1} {\bf #2}, #3 (#4)}
\font\sevenrm=cmr7
\font\fiverm=cmr5

\def\prl{\em Phys. Rev. Lett.}

\def\jgr{\em J. Geophys. Res.}
\def\grl{\em Geophys. Res. Lett.}
\def\ssr{\em Space Sci. Rev.}
\def\apj{\em Astrophys. J.}
\def\apjl{\em Astrophys. J. Lett.}
\def\apjs{\em Astrophys. J. Supp.}
\def\mnras{\em M.N.R.A.S.}
\def\phrep{\em Phys. Reports}
\def\aap{\em Astron. Astrophys.}
\def\nat{\em Nature}
\def\pss{\em Planet. Sp. Sci.} 
\def\teq#1{$\, #1\,$}
\def\dover#1#2{\hbox{${{\displaystyle#1 \vphantom{(} }\over{
   \displaystyle #2 \vphantom{(} }}$}}
\def\Pesc{P_{\hbox{\sevenrm esc}}} 
\def\valf{v_{\hbox{\fiverm A}}} 
\def\Machson{{\cal M}_{\hbox{\fiverm S}}} 
 
\def\Thetab{\Theta_{\hbox{\sevenrm B1}}}
\def\Thetabtwo{\Theta_{\hbox{\sevenrm B2}}}
{\catcode`\@=11                                                  
\gdef\SchlangeUnter#1#2{\lower2pt\vbox{\baselineskip 0pt\lineskip0pt    
\ialign{$\m@th#1\hfil##\hfil$\crcr#2\crcr\sim\crcr}}}}           
\def\gtrsim{\mathrel{\mathpalette\SchlangeUnter>}}               
\def\lesssim{\mathrel{\mathpalette\SchlangeUnter<}}    

\def\figcaption#1{\vbox{\smallskip \baselineskip10.7pt 
   \footnotesize #1\smallskip}}
\begin{document}
\vspace*{4cm}
\input{psfig.tex}
\title{DIFFUSIVE SHOCK ACCELERATION: THE FERMI MECHANISM}

\author{Matthew G. BARING \footnote[1]{Compton Fellow, USRA.
   \hskip 15pt Email: \it Baring@lheavx.gsfc.nasa.gov}}

\address{Laboratory for High Energy Astrophysics, Code 661,\\
NASA/Goddard Space Flight Center, Greenbelt, MD 20771, U.S.A.}

\maketitle\abstracts{
The mechanism of diffusive Fermi acceleration at collisionless plasma
shock waves is widely invoked in astrophysics to explain the appearance
of non-thermal particle populations in a variety of environments,
including sites of cosmic ray production, and is observed to operate at
several sites in the heliosphere.  This review outlines the principal
results from the theory of diffusive shock acceleration, focusing first
on how it produces power-law distributions in test-particle regimes,
where the shock dynamics are dominated by the thermal populations that
provide the seed particles for the acceleration process.  Then the
importance of non-linear modifications to the shock hydrodynamics by
the accelerated particles is addressed, emphasizing how these
subsequently influence non-thermal spectral formation.
}

\section{Introduction}
\label{sec:intro}

The concept of diffusive acceleration of particles in space plasmas has
been around for nearly five decades.  Fermi\cite{fermi} (1949) first
postulated that cosmic rays could be produced via diffusion between
collisions between interstellar clouds.  If such clouds have
predominantly random directions of motion, then the frequency of
``head-on'' collisions between the cosmic rays and the clouds would
exceed the rate of ``tail'' encounters, leading to a net acceleration
that is diffusive in energy.  The elegance of Fermi's idea was founded
on the fact that, when particles are confined, such a diffusive process
naturally produces power-law\cite{fermi} cosmic ray distributions,
thereby modelling the observations well.  It was subsequently realized
that shocks in space plasmas could also provide such diffusive
acceleration in an efficient manner, tapping the dissipative potential
of the flow discontinuity by transferring the shock's kinetic energy to
non-thermal populations both upstream and downstream of the shock, at
the same time as heating the downstream gas.  Thus the notion of
diffusive shock acceleration, or the {\it first-order Fermi} mechanism,
was born.

Fermi acceleration at shocks is applied to many physical systems in the
solar system, in our galaxy, and throughout the universe, primarily
because (i) of its great efficiency for generating non-thermal ions and
electrons, and (ii) since inferences of non-thermal particles abound in
astrophysics in addition to the direct observations of such populations
in the heliosphere.  The closest environment for studying such
acceleration is the Earth's bow shock, a standing shock formed by the
Earth's passage through the solar wind.  Observations of non-thermal
ions on either side of quasi-parallel portions (where the magnetic
field is normal to the shock interface) of the bow shock (e.g. Gosling
et al.\cite{gos78}; Ellison et al.\cite{emp90}) are a testament to the
efficiency of diffusive shock acceleration.  Bow shocks of other
planets and comets also exhibit particle acceleration.  Throughout the
solar system, a whole host of travelling {\it interplanetary} shocks
and forward/reverse shock pairs in {\it co-rotating interaction
regions} propagate through space, all providing unequivocal evidence
for acceleration of protons, alpha particles and other ions (e.g.
Gosling et al.\cite{gos81}; Baring et al.\cite{boef97}).  These
non-linear disturbances exemplify the complexity of the plasma and
magnetic structure of the solar wind in its journey to the outer solar
system, at which point it is commonly believed to encounter at some
distance greater than 70 AU a solar wind {\it termination shock}: this
defines the boundary between the heliosphere and the surrounding
interstellar medium.  Evidence for the existence of such a structure
is still circumstantial.

In astrophysics, the number of sites for shock acceleration burgeons.
All these are based on inferences obtained from non-thermal radiation,
given that in situ measurements are impossible.  Stellar winds (similar
to the solar wind), supernova remnants (SNRs), accreting X-ray binaries
and pulsar wind termination shocks are some examples of sources of
shock acceleration within the Milky Way.  For example, supernova
remnants commonly exhibit large ``surfaces'' of sharp intensity
variations in radio and X-ray images (e.g. see references in the
catalogue of Green\cite{dgreen95}).  Such interfaces, with a variety of
spatial morphologies, are naturally expected as the supernova ejecta in
the remnant ploughs into the surrounding interstellar medium (ISM).
Concurrent with the existence of such interfaces is non-thermal radio
synchrotron radiation in remnants, and the recent discovery of
non-thermal X-rays from remnants (e.g. Koyama et al.\cite{koy95}).
SNRs have long been thought to be the preferred site for galactic cosmic
ray generation for energies up to the so-called {\it knee} in the
spectrum at \teq{\sim 10^{15}}eV (e.g. see Lagage and
Cesarsky\cite{lg83}), where significant deviations from power-law
behaviour are observed.  The highest energy cosmic rays (\teq{\gtrsim
10^{18}}eV) are believed to be of extragalactic origin, where the large
size scales permit acceleration to energies beyond the capabilities of
galactic environments.  They could be provided by shocks associated
with jets and hot spots from radio galaxies, whose jets display a
clumpiness (knots) reminiscent of shocks (e.g. Blandford and
Eichler\cite{be87}).  There are also proponents of the idea that the
most energetic cosmic rays at \teq{\sim 10^{20}}eV may be caused by
shocks in gamma-ray bursts (Waxman\cite{wax95}; Milgrom and
Usov\cite{mu96}).  The prevalence of non-thermal X-ray emission in
Seyfert galaxies and gamma-rays from blazars promotes the notion that
the central regions of active galaxies possess shocks that efficiently
accelerate electrons and/or ions.  The underlying theme that should be
borne in mind is that wherever shocks are seen, there is evidence of
non-thermal particles:  this strongly suggests that shock acceleration
is ubiquitous in the universe.

The modern era of shock acceleration theory began with a collection of
seminal papers in 1977-78 (Krymsky\cite{krym77}; Bell\cite{bell78};
Axford, Lear and Skadron\cite{als78}; Blandford and
Ostriker\cite{bo78}) that outlined the basic properties of the
process.  Since then the field has grown substantially, with numerous
approaches being developed, each with their different attributes and
limitations.  In this review, the basic properties of shock
acceleration theory that are {\it most relevant} to astrophysical
modelling and data comparison will be reviewed, starting with a brief
outline of the principal of the Fermi mechanism.  Then the main result
of the linear or {\it test particle} theory of acceleration, namely the
emergence of canonical power-law distributions, will be derived, and a
discussion of simulation results and predictions of acceleration
efficiencies will be reviewed.  The importance of diffusion of
particles perpendicular to the ambient field for quasi-perpendicular
shocks will be discussed.  Finally the importance of non-linear
feedback between the acceleration process and the shock hydrodynamics
will be addressed, an effect that is crucial to the understanding of
many astrophysical shocks.  There are many discussions of shock
acceleration in the literature, however three principal reviews with
complementary focuses and approaches come to mind.  Drury\cite{drury83}
provides an in-depth analytic description of linear and non-linear
shocks, tailored for the specialist, while Blandford and
Eichler\cite{be87} adopt a slightly tempered analytic approach and
connect closely with astrophysical observations.  Jones and
Ellison\cite{je91} emphasize the more recent model developments 
and data/theory comparisons afforded by computer simulations.

\section{The Principal of the Fermi Mechanism}
\label{sec:Fermi}

The Fermi acceleration mechanism is always applied to so-called {\it
collisionless} shocks, i.e. those non-linear disturbances that have
energy and momentum transfer between particles mediated purely by
plasma processes, with Coulomb scattering being negligible.  Such
conditions arise in most astrophysical environments.  It is instructive
to review the principal of how the Fermi mechanism operates.  Consider
a flow defined by speeds \teq{u_1} and \teq{u_2} (\teq{< u_1}) on
different sides of the shock (see Fig.~1).  Suppose particles of speed
\teq{v_0} (in the rest frame of the shock) start off on the upstream
(\teq{u_1}) side of the shock, and diffuse around via ``collisions''
with plasma magnetic turbulence until they eventually cross the shock
and move around downstream (\teq{u_2}).  Such diffusion tends to
isotropize the angular distribution of the particles in the frame in
which the upstream plasma is at rest.  After a period in which the
particles have experienced a plasma of speed \teq{u_1}, the particles
now collide with magnetic turbulence that is associated with, or
generated by, the downstream plasma.  If this plasma in turn tends to
isotropize the particles elastically, then this test population
effectively ``sees'' a plasma moving {\it towards} it (with speed
\teq{\sim \vert u_1-u_2\vert}) upon arrival downstream.  Elementary
kinematics leads to the result that the process of quasi-isotropization
then yields a net {\it increase} in the average particle speed in the
rest frame of the shock interface.  This kinematic guarantee of an
increase in energy is akin to the gain that a photon experiences in
inverse Compton scattering collisions with relativistic electrons, an
effect that relies on photon quasi-isotropization in the electron rest
frame.

%
\vskip+0.5truecm
\centerline{\psfig{figure=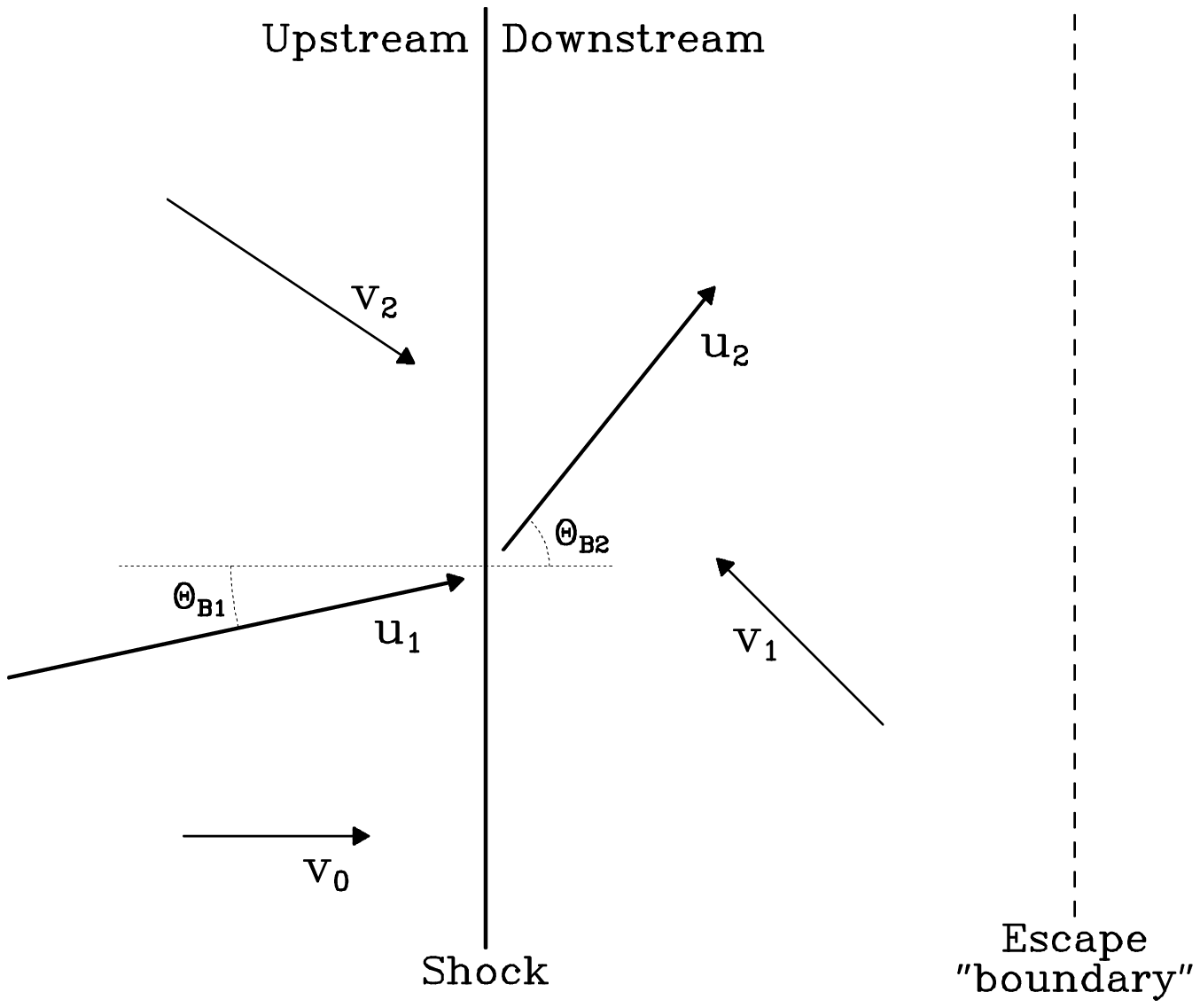,width=7.3truecm}
   \hskip 0.7truecm\psfig{figure=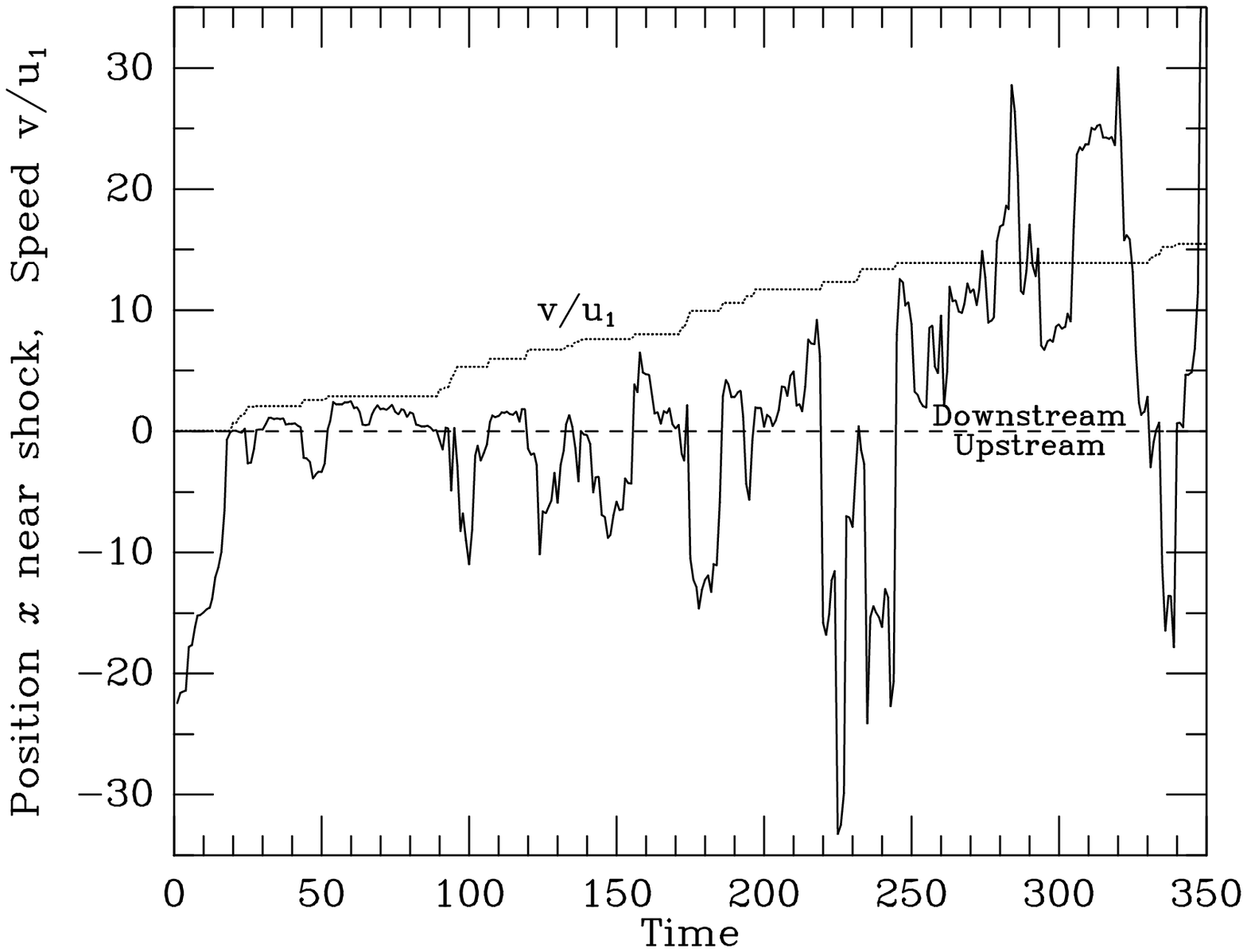,width=7.9truecm}}
\vskip-0.0truecm
\figcaption{Figure~1:  A schematic depiction (left) of particle
motion in the environs of a shock.  The plasma flow speeds are
\teq{u_1} (upstream) and \teq{u_2} (downstream), and the mean
accelerated particle speeds \teq{v_i} after \teq{i} shock crossings are
ordered according to \teq{v_0 < v_1 < v_2}, etc.  The dotted line
denotes a downstream ``escape boundary'' that is relevant to the
discussion on universal power-laws in Section~\ref{sec:testp} below.
On the right is output from a Monte Carlo simulation of first-order
Fermi acceleration (Baring et al.\cite{bej93}), illustrating how the
particle speed \teq{v} increases monotonically with time as many shock
crossings are encountered, until it escapes the shock (dotted line).
Notice that the diffusion is on larger scales for higher particle
speeds.}

\newpage

Eventually, some of the particles will return to the upstream side of
the shock, and these will see the upstream plasma moving, on average,
towards them.  Hence the process is repeated, and diffusion that leads
to significant isotropization in the local plasma frame will yield
particle speeds that are on average higher than those previously
obtained on the downstream side of the shock.  Sequential diffusion
back and forth across the shock always leads to increases in particle
speed, so that many shock crossing cycles afford significant
acceleration.  This is the principal of the {\it first-order} Fermi
mechanism, where energy gains are always positive, and spatial
diffusion allows the shock to dissipate its energy via its associated
turbulence to a non-thermal particle population.  This situation is
depicted in Fig.~1, both schematically, and also in an example where an
actual Monte Carlo simulation\cite{bej93} was used to demonstrate the
coupling between shock crossings and energy increase.  The velocity
increase, for non-relativistic particles, is proportional to \teq{\sim
\vert u_1-u_2\vert}, a feature that is evident in Fig.~1, so that many
cycles are required for particles to achieve high energies.  For a
simple visualization, the Fermi mechanism can be likened to specular
reflection between two converging mirrors, for example a ball bouncing
elastically back and forth between two walls that approach each other.

Before moving to a discussion of aspects of test-particle acceleration,
a brief summary of approaches to shock acceleration theory is
desirable.  Early in the 1980s, {\it two-fluid} models (see
Drury's\cite{drury83} review) were developed, treating the cosmic rays
and thermal gas as separate entities, and exploring the hydrodynamics
of shocked flows subject to non-linear shock acceleration.  These are
extremely useful for time-dependent applications, but contain little
spectral information.  Differential equations that describe the
convective and diffusive parts of particle motion have been used in a
variety of situations with a variety of assumptions; they handle
time-dependence and spectral information well, but have difficulty
injecting particles from the thermal population.  Plasma
codes\cite{je91} compute particle motions and field structure using the
Newton-Lorentz and Maxwell's equations, and hence are physically the
most detailed approach to acceleration theory.  They come in full and
hybrid (where electrons are treated as a background fluid) varieties,
are great for studying wave properties, but have a severely CPU-limited
spectral dynamic range.  Finally, kinetic Monte Carlo
simulations\cite{je91}, of which several examples are presented here,
describe the convective and diffusive particle motions using a
prescribed scattering or diffusion law.  They self-consistently inject
particles from thermal energies, have excellent spectral information
and dynamic range, and hence are ideal for comparison with experimental
data.  The Monte Carlo technique is presently limited to
time-independent situations.

\section{Acceleration in Test-Particle Regimes}
\label{sec:testp}

The ions accelerated by the Fermi mechanism are called test particles
when they do not have sufficient pressure (i.e. energy density) to
influence the hydrodynamics of the shocked plasma, and hence ``go along
for the ride.''  The description of the shock acceleration mechanism is
then linear and therefore comparatively simple.  Moreover, when the
test particle speeds far exceed that of the shock, or equivalently
\teq{u_1} and \teq{u_2}, the system possesses no velocity scale (not
true of non-linear acceleration) so that the natural solution for the
distribution function of the non-thermal particles (also frequently
referred to as the cosmic ray component) is a power-law distribution.
This is the foremost property of the test-particle, or linear,
acceleration regime, and it is instructive to derive the ``canonical''
power-law index.

Let the distribution function for particles of speed \teq{v} and
momentum \teq{p=mv} in some frame of reference (say the shock frame) be
\teq{f(p)}, and the cumulative distribution between \teq{p} and
infinity be \teq{{\cal F}(p)}.  Assume that \teq{v\gg u_1} and that the
particles are isotropic; under these conditions, the choice of
reference frame among the upstream and downstream plasma frames or the
shock frame is immaterial---isotropy is attained in all three.  Then it
is possible to write down the differential equation governing
\teq{{\cal F}(p)}:  in a steady-state scenario, the system is described
by
\begin{equation}
   0\;\equiv\; t_{cyc}\,\dover{d{\cal F}}{dt}\; =\;
   -\langle\Delta p\rangle\,\dover{\partial {\cal F}}{\partial p}
   -\Pesc {\cal F}\quad , \quad {\cal F}(p)\, =\,
   \int_p^{\infty} f(p_1)\, dp_1\quad .
 \label{eq:dfdt}
\end{equation}
Here \teq{t_{cyc}} is the mean time for a cycle of {\it two}
transmissions through the shock, e.g. upstream to downstream and then
back upstream again, and \teq{\langle\Delta p\rangle} is the mean
(positive) net momentum gain in a cycle.  Hence the \teq{
-\langle\Delta p\rangle\, \partial {\cal F}/\partial p} term represents
the supply of particles to higher energies via the acceleration
process, and involves a derivative with respect to \teq{p} because the
momentum gains are approximately differential: \teq{\langle\Delta
p\rangle\sim m\vert u_1 -u_2\vert\ll p}.  The negative sign appears in
this term due to the use of a cumulative distribution.  The remaining
term represents a loss of particles from each cycle (with probability
\teq{\Pesc}) beyond some downstream reference boundary that is depicted
in Fig.~1.  The existence of such a loss is a consequence of the net
sense of the plasma flow in the downstream direction; when \teq{v\gg
u_1} then the spatial diffusion of the particles back and forth across
this boundary dominates convection from the flow and the probability of
escape is small (\teq{\Pesc\ll 1}).  This escape term is proportional
to \teq{{\cal F}(v)} since particles that are lost at a particular
speed cannot contribute further to the acceleration process.

The form of Eq.~(\ref{eq:dfdt}) corresponds to the first-order Fermi
process, so named because the acceleration is first-order in velocity
differentials with the equation having only friction terms in \teq{p}.
Second-order Fermi acceleration, for which there is diffusion in
\teq{p}, can also occur and yield power-law distributions.  It is
generally a minor contribution unless the shock speed is almost as low
as the Alfv\'en speed \teq{\valf =B/\sqrt{4\pi\rho}}, which defines the
``speed'' of the magnetic disturbances in the plasma that effect
particle scattering and diffusion.  Hence Alfv\'en speeds of the order
of \teq{u_1} lead to extra terms in the Fokker-Planck expansion that
has its friction term appearing in Eq.~(\ref{eq:dfdt}).  Solutions to
this equation are easily obtained.  For isotropy of particles, the
probability of downstream escape of particles with \teq{v\gg u_1} from
a cycle is \teq{\Pesc\approx 4u_1/v} for {\it plane-parallel} shocks
where the field is normal to the discontinuity interface.  The average
momentum gain in a cycle for isotropic populations is \teq{\langle
\Delta p\rangle\approx 4u_1 (u_1-u_2)p/(3u_2v)}.  Hence, defining the
shock {\it compression ratio} as \teq{r=u_1/u_2}, Eq.~(\ref{eq:dfdt})
admits power-law solutions (for \teq{u_1\ll c}):
\begin{equation}
   f(v)\;\propto\; p^{-\sigma}\;\;, \quad \sigma\, =\, 1+
   \dover{p\Pesc}{\langle \Delta p\rangle}\, =\, \dover{r+2}{r-1}\quad ,
   \quad r\, =\,\dover{u_1}{u_2}\quad .
 \label{eq:powerlaw}
\end{equation}
This is the canonical or universal power-law distribution that is
frequently invoked in shock acceleration applications to astrophysics,
and is valid both for non-relativistic and relativistic particles.  The
power-law index \teq{\sigma} depends only on the compression ratio
\teq{r}, an elegant feature of the first-order Fermi mechanism, and
non-relativistic plasma shocks (of adiabatic index \teq{\gamma =5/3})
with large sonic Mach numbers (so-called {\it strong} shocks) have
\teq{r=4} and therefore \teq{\sigma =2}.  If the plasma speeds
\teq{u_{1,2}} used in computing the compression ratio are interpreted
as flow velocity components along the shock normal, then the result in
Eq.~(\ref{eq:powerlaw}) extends to {\it oblique} shocks (e.g.
Drury\cite{drury83}), where the mean magnetic field {\bf B} makes
significant angles with the shock normal (see Fig.~1).  Furthermore,
the canonical power-law is applicable regardless of the relative
contributions of particle diffusion along and orthogonal to {\bf B}
(Jones\cite{jones94}).  Alternative derivations of the universal
power-law are presented by Bell\cite{bell78}, Drury\cite{drury83},
Blandford and Eichler\cite{be87} and Jones and Ellison\cite{je91}.
Note that while second-order Fermi acceleration also yields power-laws,
there is no coupling between \teq{\sigma} and the compression ratio
\teq{r}, primarily because \teq{\sigma} depends on two independent
parameters: the speed of the scatterers (i.e. magnetic turbulence) in
the flow frame, and the residence time in the acceleration region.
Hence there is {\bf no universality of the power-law} for 2nd-order
acceleration, underlining the beauty of the first-order process.

In the shock acceleration mechanism, the canonical power-laws are
attained only at high energies, so that a more complex (i.e.
non-analytic) description becomes necessary when \teq{v\sim u_1}.
Measurements of shock-associated non-thermal ions in the heliosphere
clearly indicate that low energy non-thermal ions are drawn directly
from the thermal component of the plasma.  In particular, it seems that
all ions are subject to the same diffusive behaviour, with no peculiar
situation arising at either low or high energies.  Hence, a coherent
treatment of Fermi acceleration must give similar weight to thermal and
superthermal particles; this is a nice feature of some of the computer
simulations reviewed by Jones and Ellison\cite{je91}.  Such simulations
provide insight into the relative abundances of thermal and non-thermal
ion components; these reflect the inherent efficiency of the
acceleration process.  Therefore, we turn our attention now to survey
results from simulations that address salient properties of linear
shock acceleration theory, such as the dependence of the shape of the
distribution and the acceleration efficiency on shock obliquity,
particular plasma characteristics and the description of scattering.

%
\vskip+0.5truecm
\centerline{\psfig{figure=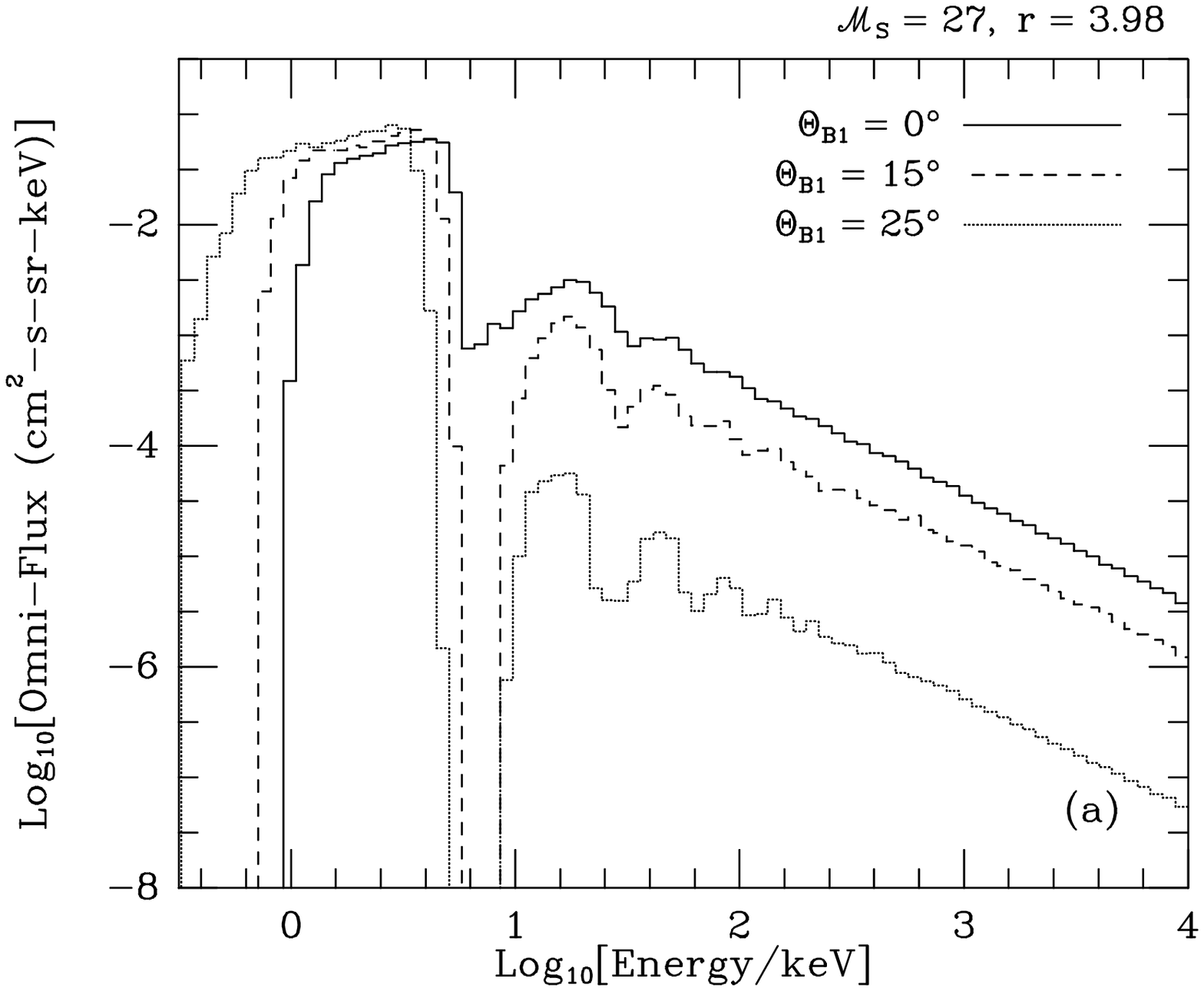,width=7.85truecm}
   \hskip 0.2truecm\psfig{figure=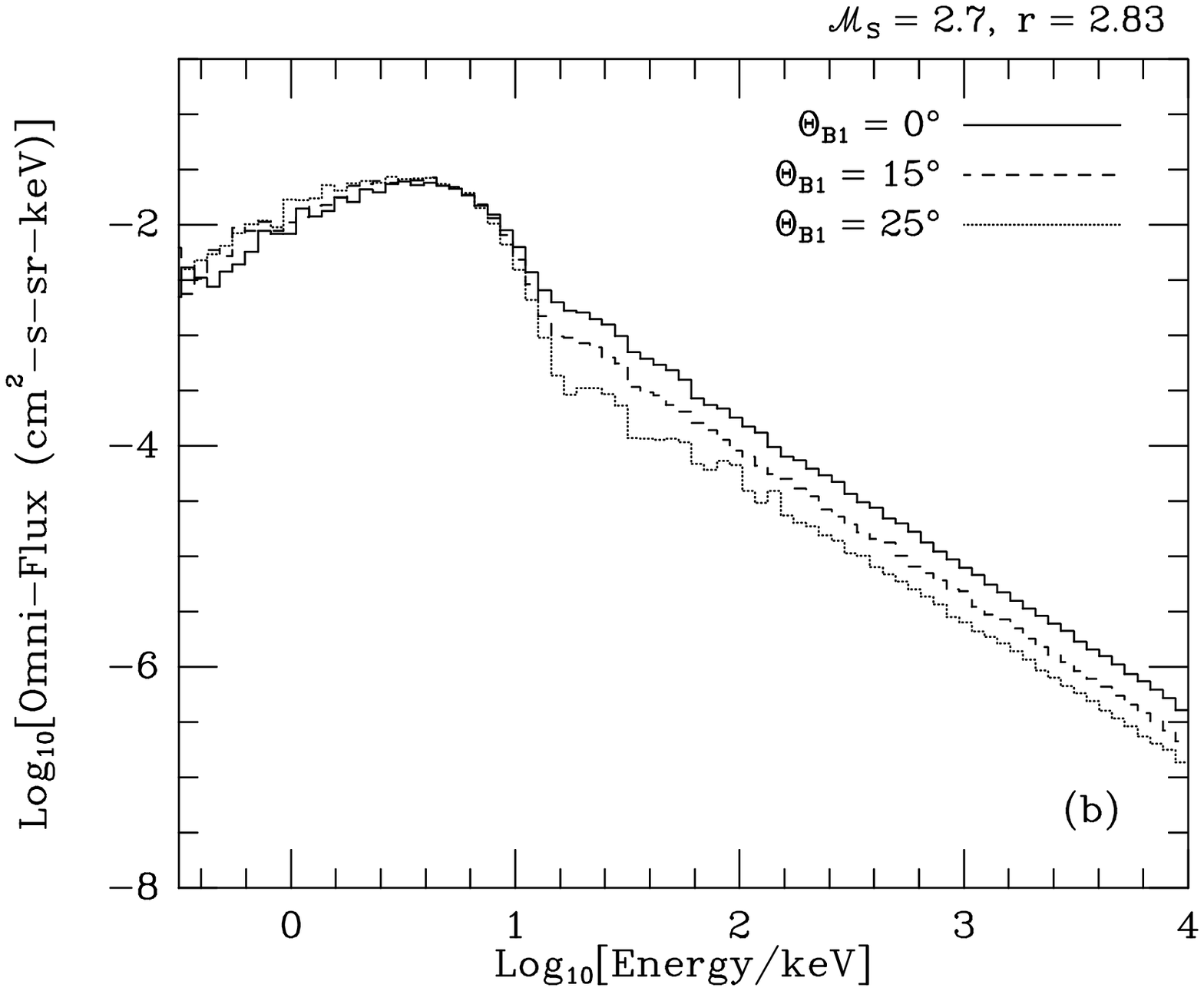,width=7.85truecm}}
\vskip-0.0truecm
\figcaption{Figure~2:  Omni-directional flux distributions for
different shock obliquities, normalized to unity and measured
downstream of (a) high and (b) low sonic Mach number shocks of speed
1000 km/sec, as obtained\cite{bej93,bej94} from kinematic Monte Carlo
simulations of Fermi acceleration.  Particles were injected into the
upstream region, with temperatures \teq{T_1=10^5}K (\teq{\Machson =27})
and \teq{T_1=10^7}K (\teq{\Machson =2.7}), and were allowed to escape
from the shock either far downstream or when their energies exceeded
\teq{10^4}keV.  A dramatic reduction in injection efficiency with
increasing obliquity is evident in the high Mach number case (a) with
the normalization dropping by about a factor of 100 between
\teq{\Thetab=0^\circ} and \teq{25^\circ}.  The power-law indices
closely approximate those in Eq.~(\ref{eq:powerlaw}). }

Although the power-law index of the accelerated particles is
independent of shock obliquity, the acceleration efficiency is {\it
strongly} dependent on the angle \teq{\Thetab} that the upstream
magnetic field makes to the shock normal.  This is illustrated in
Fig.~2, where ``spectral'' results from a kinematic Monte Carlo
approach to shock acceleration (Baring et al.\cite{bej93}) are shown.
The distributions, obtained in a region just downstream of the shock,
exhibit prominent thermal components at low energies, and canonical
power-law tails at high energies, with structure in between (for
\teq{u_1\lesssim v\lesssim 10u_1}).  The prominence of this structure
originates in the simplistic nature of the particle trajectories
assumed in these calculations, and is largely smeared out by more
detailed approaches (e.g. Ellison et al.\cite{ebj95}) that include
particle gyromotions.  Notwithstanding, the structure signifies
successive shock crossings, ordered according to increasing energy, and
merges smoothly into the power-law tail when \teq{v\gg u_1}.  In the
left hand panel, where the upstream plasma is very cold (and the sonic
Mach number \teq{\Machson\approx u_1\sqrt{m_p/(\gamma kT_1)}} is large)
the thermal peaks are not truly Maxwellian, but rather are defined by
the kinematics of the simulation:  a more complicated picture of shock
dissipation and heating would yield almost Maxwellian shapes (e.g. see
results from detailed plasma simulations in Giacalone et
al.\cite{gbse92}, Scholer et al.\cite{stk92}).  In the right hand
panel, the higher temperature (lower \teq{\Machson}) smears out the
structure.  Note that for lower \teq{\Machson}, the hotter upstream
plasma produces weaker shocks (lower \teq{r}) and therefore steeper
distributions.  The normalization of the non-thermal component relative
to the thermal population describes the overall efficiency of
acceleration.  This is seen to drop dramatically for large Mach numbers
when significant obliquities \teq{\Thetab} are realized.  The reason
for this (Baring et al.\cite{bej93,bej94}) is that as the obliquity
increases, the mean increment in the particle speed \teq{\langle \Delta
v\rangle} upon first crossing of the shock declines, and turns out to
approach zero in the limit of {\it quasi-perpendicular} (i.e.
\teq{\Thetab\approx 90^\circ}) shocks.  When the component of this
increment along the mean field direction drops below \teq{u_2},
particles cannot return to the upstream side of the shock so that
injection into the Fermi process fails.  This occurs\cite{bej94} when
\teq{\Thetab\gtrsim 30^\circ} for cold upstream plasmas and at
\teq{\Thetab\gtrsim 55^\circ} for the warm plasma case in right hand
panel of Fig.~2.

This termination of injection poses a serious problem for highly
oblique and quasi-perpendic-ular shocks.  These disturbances abound in
the heliosphere as travelling interplanetary shocks and discontinuities
associated with co-rotating interaction (CIR) regions.  Due to the
tightness of the winding of the spiral heliospheric field in the solar
wind, most interplanetary and CIR shocks are highly oblique.  Yet they
are observed to be prolific accelerators of ions (for recent Ulysses
observations see, e.g., Ogilvie, et al.\cite{ogil93}; Gloeckler, et
al.\cite{gloe95}; Baring et al.\cite{boef97}).  Therefore, the theory
leading to the distributions in Figure~2 lacks an ingredient that is
key to the acceleration process in highly oblique shocks --- this
ingredient is {\it cross-field diffusion}.  The simulation results in
Figure~2 included only a description of particle diffusion along the
local magnetic field lines.  Clearly turbulent plasmas provide
diffusion across the mean direction of {\bf B}, as is deduced from
spacecraft observations of interplanetary magnetic field power spectra
(e.g. Forman et al.\cite{form74}; Valdes-Garcia et al.\cite{vqm92}).
To add this to Monte Carlo simulations effectively requires
specification of a {\it spatial diffusion coefficient} perpendicular to
the mean local (i.e. upstream or downstream) field,
\teq{\kappa_{\perp}}.  Using the standard kinetic theory result for
field turbulence (e.g.  Axford\cite{axf65}; Forman et
al.\cite{form74}), this can be related to the diffusion coefficient
along the field, \teq{\kappa_{\parallel}}, via
\begin{equation}
\kappa_{\perp}\; =\;\dover{\kappa_{\parallel}}{1 +(\lambda/r_g )^2}\quad .
 \label{eq:kappas}
\end{equation}
Here \teq{\lambda} is the mean free path for particle scattering (i.e.
diffusion) along the field, which is related to the spatial diffusion
coefficient by \teq{\kappa_{\parallel}=\lambda v/3}.  Also, \teq{r_g=
pc/(qB)} is the gyroradius of a particle of momentum \teq{p} and charge
\teq{q}.  In the limit of small mean free paths, \teq{\lambda\sim r_g},
the diffusion satisfies \teq{\kappa_{\perp}\sim \kappa_{\parallel}} and
is quasi-isotropic.  This defines the {\it Bohm} diffusion
limit where the field turbulence is strong and no direction is
preferred for particle diffusion; in this limit the Fermi acceleration
of ions should be more or less equivalent for quasi-parallel
(\teq{\Thetab\sim 0^\circ}) and quasi-perpendicular (\teq{\Thetab\sim
90^\circ}) shocks.  Values of \teq{\lambda\lesssim r_g} are
physically unrealistic.

The inclusion of cross-field diffusion short-circuits the termination
of injection, provided that it can return to the shock particles that
have entered the downstream region for the first time.  Since particles
diffuse of the order of their gyroradius \teq{r_g} perpendicular to the
field every time they are ``scattered'' along the field (i.e. on the
scale of \teq{\lambda}), simple geometry considerations imply that the
criterion for the rejuvenation of injection is that the ratio
\teq{\lambda /r_g} be small enough, and hence that the ratio of
\teq{\kappa_{\perp}/\kappa_{\parallel}} be sufficiently large [but
still less than unity: see Eq.~(\ref{eq:kappas})].  The critical value
of \teq{\lambda /r_g} is dependent on the downstream field angle
\teq{\Thetabtwo} to the shock normal in a non-trivial way.  Physically
meaningful values of \teq{\lambda /r_g} can always be found to effect
injection and efficient acceleration in highly oblique
shocks\cite{boef97,ebj95}.  The survey work of Ellison et
al.\cite{ebj95} found that for oblique shocks the efficiency of
non-thermal ion generation declined rapidly with increasing Mach
numbers and increasing values of \teq{\lambda /r_g}, and further that
unless the turbulence forced the situation very close (i.e.
\teq{\lambda /r_g\lesssim 2}) to the Bohm diffusion limit, there was a
significant reduction in acceleration efficiency with increasing
obliquity.  Jokipii\cite{jok87} observed that large values of
\teq{\lambda /r_g} reduced the timescale for acceleration in
quasi-perpendicular shocks, and identified this property as a means to
ease problems\cite{lg83} with accelerating cosmic rays all the way up
to the ``knee'' using supernova remnants.  Ellison et al.\cite{ebj95}
highlighted the trade-off that then arises in oblique shocks: those
that accelerate quickly are inefficient, and those shocks that produce
cosmic rays copiously take longer to do so.  This has a significant
impact on models of cosmic ray production.

\newpage

The appropriateness of small values of \teq{\lambda /r_g} for highly
oblique interplanetary shocks was affirmed by Baring et
al.\cite{boef97}, who obtained impressive fits to Ulysses distribution
data for two shocks less than about 3 AU from the sun, using a Monte
Carlo approach\cite{bej93,ebj95}.  Distributions were derived using
upstream solar wind properties as input for the simulation.  Such
accurate fits, obtained for ion speeds up to around 4--5 times the
solar wind speed, were possible only for values \teq{\lambda /r_g
\lesssim 3} in shocks with \teq{\Thetab\gtrsim 60^\circ}.  Such low
values are comparable to those deduced from spacecraft observations of
interplanetary magnetic field turbulence\cite{form74,vqm92}.  Kang and
Jones\cite{kj97} obtained similar fits in regimes of strong turbulence
using an alternative theoretical approach, the numerical solution of
the convection-diffusion differential equation for particle transport
in shocks.  An elegant aspect of the work of Baring et al.\cite{boef97}
was the ability to simultaneously model proton and alpha-particle data;
from this success, there is a strong suggestion that the particles
interact with the field turbulence in a more-or-less elastic manner.
Such theory/data comparisons are yet to be fully extended to shocks a
little more distant from the sun; these possess the added complication
of significant pick-up ion components\cite{boef97,gloe95}.  Note that
interplanetary shocks have low Mach numbers, and are therefore weak;
they are thus comparatively inefficient accelerators, so that using
test-particle theories to model them is quite fitting.

The discussion so far has focused on non-relativistic shocks, for which
there is an abundance of observational data in the heliosphere and
applications in astrophysics.  It is appropriate to comment briefly on
results for relativistic shocks (where \teq{u_1\gtrsim 0.1c}), which
are less comprehensively studied, yet may be quite important for active
galaxies.  Relativistic gases have adiabatic indices of \teq{\gamma
=4/3}, and hence are more compressible than their non-relativistic
counterparts: they can generate compression ratios as large as
\teq{r=7} in linear shock applications.  Shocks with \teq{u_1\gtrsim
0.1c} produce power-law distributions of relativistic particles, yet
Eq.~(\ref{eq:powerlaw}) cannot be used to predict their power-law index
primarily because the assumption of isotropy must be relinquished.
Relativistic shock fronts move sufficiently fast that even
ultrarelativistic particles have difficulty catching them.  Hence
escape from the shock is greater and particle isotropy is impossible to
achieve.  At the same time, the energy boosts that ions receive in
diffusion back and forth across the shock are enhanced relative to
non-relativistic shock cases.  The net effect is that for
\teq{f(p)\propto p^{-\sigma}}, the spectral index \teq{\sigma}
decreases with increasing shock speed \teq{u_1}.  This behaviour was
identified in the pitch-angle diffusion (i.e. imposing incremental
changes in particle direction) transport equation analysis of Kirk and
Schneider\cite{ks87} and the large angle scattering kinematic Monte
Carlo simulation of Ellison et al.\cite{ejr90}.  A comparison of their
results\cite{ejr90} reveals that the type of diffusion, i.e. pitch
angle or large angle, strongly affects the population anisotropy and
therefore is crucial to the determination of the power-law index.

The absence of the mention of electrons in the above discussions is
striking.  This is largely because the injection of electrons into the
Fermi process is at present poorly understand.  While protons at all
energies from thermal upwards can resonate with Alfv\'en waves, it is
not clear what waves thermal electrons can interact with to effect
diffusion.  Above around 20 MeV, electrons resonate with Alfv\'en waves
and can actively participate in the diffusive acceleration process
producing results\cite{re92} similar to those addressed above.
However, between around 20 MeV and 0.5 MeV electrons can ``scatter''
off {\it whistler} waves (e.g. Levinson\cite{lev92}).  Yet below this
energy the picture is particularly unclear, and is complicated by the
fact that whistlers can be strongly damped in hot plasmas; Alfv\'en
modes usually escape this fate.  Given ubiquitous observations of
non-thermal radiation in astrophysics, as most non-thermal electronic
radiative processes are so much more efficient than ionic ones, the
inference of non-thermal (and predominantly relativistic) electron
populations in numerous sources seems robust.  Hence, if shock
acceleration is assumed to be the origin such electrons, it remains to
solve the problem of their injection into the Fermi mechanism.  The
role of electrons is discussed in the accompanying paper (Baring, this
volume) that reviews applications of shock acceleration to gamma-ray
production in young supernova remnants.

\newpage

\section{Non-Linear Modifications: a Brief Look}
\label{sec:nonlinear}

Linear test-particle models of shock acceleration are valid provided
that the accelerated population does not alter the flow hydrodynamics.
For strong, non-relativistic shocks \teq{r=4}, Eq.~(\ref{eq:powerlaw})
indicates a \teq{p^{-2}} distribution, which provides a divergent
cosmic ray pressure or energy density if the momenta are permitted
to extend to infinity.  Consequently, if the maximum momentum achieved in
a particular shock environment is large enough, the accelerated
population dominates the gas pressure and can therefore modify or
dominate the flow dynamics.  This maximum \teq{p} is generally determined
by obvious natural limits such as the diffusion length of particles
(greater than their gyroradii) being less than the physical size of the
astrophysical system, beyond which particle escape ensues.  The standard
approach for describing such non-linear modifications to the flow is
using the {\it Rankine-Hugoniot} relations for conservation of particle
number, momentum, and energy fluxes.  These relations form the central
theme of the earliest approach to non-linear shock acceleration theory,
the {\it two-fluid} (i.e. thermal gas plus cosmic rays) model\cite{dv81},
reviewed comprehensively by Drury\cite{drury83}.  The energy flux is
the highest order momentum moment of the distribution: hence necessity
of a non-linear treatment is contingent upon the pressure of the cosmic
rays being a sizeable fraction of that of the thermal gas.  Generally,
the consideration of non-linear effects is imperative\cite{ee84} for
shocks with \teq{r\gtrsim 3.4}.

The non-linearity is manifested through the feedback of the particles
on the spatial profile of the flow velocity\cite{drury83}, which in
turn determines the shape of the distribution.  The accelerated
population presses on the upstream plasma and slows it down before the
discontinuity is reached.  An upstream {\it precursor} forms, in which
the flow speed is monotonically decreasing.  At the same time, the
cosmic rays press on the downstream gas, slowing it down also.  The net
effect that usually emerges is one where the overall compression ratio,
from far upstream to far downstream of the discontinuity, actually {\bf
exceeds that of the test-particle scenario}.  This phenomenon was
identified by Eichler\cite{eich84}, and Ellison and Eichler\cite{ee84},
and arises because of the possibility of particle escape at high energies;
such a drain of energy and momentum flux must be compensated by ramping
up the overall compression ratio to conserve the fluxes.  It is
illustrated and discussed in more detail in the accompanying paper by
Baring in this volume (see also Jones and Ellison\cite{je91}).  The
highest energy particles generally have longer diffusion lengths\cite{gbse93}
(unless \teq{\kappa} is chosen to be independent of energy\cite{das82},
an improbable situation), so that they sample a broader portion of the
flow velocity profile, and hence larger compressions ratios.
Consequently, these particles have a flatter power-law index than those
at lower energies, thereby dominating the pressure in a non-linear
fashion.  The resulting upward spectral curvature\cite{ee84,eich84} is
the trademark of non-linear acceleration.  Such spectral properties can
be probed by Monte Carlo and plasma simulations, but cannot easily be
explored using the two-fluid technique.

It remains to point out that such non-linear considerations are
essential in certain real applications, even though they are seldom
implemented in astrophysical models.  The seminal work in this regard
was performed by Ellison, M\"obius and Paschmann\cite{emp90}, who used
the Monte Carlo technique to fit AMPTE data just upstream and downstream
of quasi-parallel portions of the Earth's bow shock.  The fit was
impressive for thermal and non-thermal protons, and non-thermal
He$^{++}$ and a mixture of C,N and O ions (for which no thermal
distributions could be measured).  Furthermore, it {\it required} the
implementation of non-linear modifications in this strong shock, and
test-particle models were excluded by the data.  This comparison has
since been reproduced by other techniques, namely using hybrid plasma
simulations\cite{gbse92,ts91} and solutions to the convection-diffusion
transport equation\cite{kj95}.  At the same time, the importance of
non-linear considerations for electron acceleration have been outlined
by Reynolds and Ellison\cite{re92}, including the suggestion of an
inference of spectral curvature from radio data for Kepler's supernova
remnant.  The role of non-linear shock modifications on electron
acceleration is examined at greater length in accompanying paper
(Baring, this volume) on SNR applications.  The necessity of non-linear
considerations to supernova remnant modelling is underlined in the
discussions there, augmenting the brevity of this exposition.  To
conclude, {\it non-linear} shock acceleration theory is central to the
modelling of the Earth's bow shock and SNRs, and may also be extremely
relevant to relativistic shock applications, territory that is yet to
be explored.

\vskip 9pt\noindent {\bf Acknowledgments:}
I thank my current collaborators Don Ellison, Frank Jones and Keith
Ogilvie for many productive discussions on shock acceleration theory
and observational data, and John Kirk for introducing me to the field
a while back.


\end{document}